\documentclass[preprint,12pt]{elsarticle}

\usepackage{amssymb}

\journal{Physica A}

\usepackage{caption}
\usepackage{subcaption}
\usepackage{amsmath}
\usepackage{mathtools}
\usepackage{tikz}
\usetikzlibrary{decorations.pathmorphing}
\usetikzlibrary{arrows.meta}
\usetikzlibrary{shapes,backgrounds}
\usetikzlibrary{positioning}
\usetikzlibrary{patterns}
\usepackage[utf8]{inputenc}
\makeatletter
\def\ps@pprintTitle{%
 \let\@oddhead\@empty
 \let\@evenhead\@empty
 \def\@oddfoot{}%
 \let\@evenfoot\@oddfoot}
\makeatother
\biboptions{sort&compress}

\begin{document}

\begin{frontmatter}



\title{Bias induced drift and trapping on random combs and the Bethe lattice: Fluctuation regime and first order phase transitions}


\author[inst1,inst2]{Jesal D Kotak}
\ead{jesaldkotak@gmail.com}
\author[inst1]{Mustansir Barma}
\ead{barma23@gmail.com}

\affiliation[inst1]{organization={TIFR Centre for Interdisciplinary Sciences, Tata Institute of Fundamental Research},
            addressline={Gopanpally}, 
            city={Hyderabad},
            postcode={500046}, 
            country={India}}

\affiliation[inst2]{organization={Indian Institute of Space Science and Technology},
            addressline={Valiamala}, 
            city={Thiruvananthapuram},
            postcode={695547}, 
            country={India}}

\begin{abstract}
We study the competition between field-induced transport and trapping in a disordered medium by studying biased random walks on random combs and the bond-diluted Bethe lattice above the percolation threshold. While it is known that the drift velocity vanishes above a critical threshold, here our focus is on fluctuations, characterized by the variance of the transit times. On the random comb, the variance is calculated exactly for a given realization of disorder using a ‘forward transport’ limit which prohibits backward movement along the backbone but allows an arbitrary number of excursions into random-length branches. The disorder-averaged variance diverges at an earlier threshold of the bias, implying a regime of anomalous fluctuations, although the velocity is nonzero. Our results are verified numerically using a Monte Carlo procedure that is adapted to account for ultra-slow returns from long branches. On the Bethe lattice, we derive an upper bound for the critical threshold bias for anomalous fluctuations of the mean transit time averaged over disorder realizations. Finally, as for the passage to the vanishing velocity regime, it is shown that the transition to the anomalous fluctuation regime can change from continuous to first order depending on the distribution of branch lengths.
\end{abstract}




\end{frontmatter}



\section{Introduction} \label{sec:introduction}
Transport in disordered media is important in many settings \citep{AlexanderBernasconi,  HavlinAvraham, BOUCHAUD1990127} ranging from conduction in disorder-doped semiconductors \cite{BottgerBryskin} and hydrodynamic dispersion in porous media \cite{BOUCHAUD1990127} to the diffusion of molecules in chromatography \citep{Fischer2,DeVos2016}. Especially interesting effects arise in the presence of a biasing field, which drives the particles preferentially in one direction, inducing a drift of the carriers. While a low value of bias induces a drift velocity $v_{drift}$ proportional to the bias \citep{ohtsuki1984,odagaki1981,Pury2002}, larger values lead to $v_{drift}$ becoming non-monotonic because of local traps which are easy to reach but difficult to get out of. These traps correspond to local minima of the potential energy; passage out of each trap typically involves activation, a very slow process. Various aspects of the competition between drift and trapping, both induced by the biasing field, and their effect on macroscopic transport properties have been studied in \citep{MBDDhar1983,Derrida1983,Dhar_1984,White_1984,POTTIER19951,BALAKRISHNAN19951,CaceresPRE62,Demaerel_2018}.

A good first model of a disordered medium is provided by the infinite spanning cluster in the percolation problem, which forms above a critical value $p_c$ of the fraction $p$ of randomly occupied bonds. Considering the motion of a biased random walker on this structure, it was argued that the macroscopic drift velocity $v_{drift}$ is a non-monotonic function of bias, and in fact vanishes above a critical value of the bias \citep{MBDDhar1983}, later confirmed by Monte Carlo simulations \citep{dharStauffer1998,Kirsch1998,Stauffer1999,CaceresPRE62}. Two types of traps were identified, namely dead-end branches in the direction of the field, and backbends in infinite paths along the backbone of the infinite cluster. Simple models were introduced to study the effect of the two types of traps - the random comb (RC) to model branches of random lengths, and a 1-D model with random reversals (RR) of the easy direction, to model backbends along the backbone, \cite{MBDDhar1983} (Fig. \ref{fig:RC-RR}). In both RC and RR models it was found that $v_{drift} = 0 $ beyond a critical value of the bias \cite{MBDDhar1983,Derrida1983,White_1984,BALAKRISHNAN19951,POTTIER19951}. Moreover, in the RR model, the diffusion constant was shown to diverge at a smaller critical value of bias, indicating the presence of a regime with anomalously large spreads of the profile in a range where $v_{drift}$ is non-zero. \cite{Derrida1983,aslangul:jpa-00210967,Aslangul1989}.

\begin{figure}[!htb]
    \centering
\begin{tikzpicture}[scale = 0.8]

\begin{scope}
    \draw (0,0) -- (5,-5);
      \draw[fill] (0,0) circle [radius=0.03];
      \draw[fill] (0.5,-0.5) circle [radius=0.03];
      \draw[fill] (1,-1) circle [radius=0.03];
      \draw[fill] (1.5,-1.5) circle [radius=0.03];
      \draw[fill] (2,-2) circle [radius=0.03];
      \draw[fill] (2.5,-2.5) circle [radius=0.03];
      \draw[fill] (3,-3) circle [radius=0.03];
      \draw[fill] (3.5,-3.5) circle [radius=0.03];
      \draw[fill] (4,-4) circle [radius=0.03];
      \draw[fill] (4.5,-4.5) circle [radius=0.03];
      \draw[fill] (5,-5) circle [radius=0.03];
      
      \draw (0,0) -- (-1,-1);
      \draw (0.5,-0.5) -- (-0.5+0.5,-0.5-0.5);
      \draw (1,-1) -- (-1+1,-1-1);
      \draw (2,-2) -- (-0.5+2,-0.5-2);
      \draw (2.5,-2.5) -- (-2+2.5,-2-2.5);
      \draw (3.5,-3.5) -- (-1+3.5,-1-3.5);
      \draw (4.5,-4.5) -- (-0.5+4.5,-0.5-4.5);
      
      \draw[fill] (-0.5,-0.5) circle [radius=0.03];
      \draw[fill] (-1,-1) circle [radius=0.03];
      \draw[fill] (0,-1) circle [radius=0.03];
      \draw[fill] (0.5,-1.5) circle [radius=0.03];
      \draw[fill] (0,-2) circle [radius=0.03];
      \draw[fill] (1.5,-2.5) circle [radius=0.03];
      \draw[fill] (2,-3) circle [radius=0.03];
      \draw[fill] (1.5,-3.5) circle [radius=0.03];
      \draw[fill] (1,-4) circle [radius=0.03];
      \draw[fill] (0.5,-4.5) circle [radius=0.03];
      \draw[fill] (3,-4) circle [radius=0.03];
      \draw[fill] (2.5,-4.5) circle [radius=0.03];
      \draw[fill] (4,-5) circle [radius=0.03];
      \node [below] at (2.5,-5.5) {\parbox{0.3\linewidth}{\caption*{(a) Random comb (RC)}\label{fig:RC1}}};
\end{scope}

\begin{scope}[xshift=8cm]
  \draw (0,0) -- (0.5,-0.5);
      \draw (0,-1) -- (0.5,-0.5);
      \draw (0,-1) -- (0.5,-1.5);
      \draw (1.5,-0.5) -- (0.5,-1.5);
      \draw (1.5,-0.5) -- (2,-1);
      \draw (1,-2) -- (2,-1);
      \draw (1,-2) -- (1.5,-2.5);
      \draw (2,-2) -- (1.5,-2.5);
      \draw (2,-2) -- (2.5,-2.5);
      \draw (2,-3) -- (2.5,-2.5);
      \draw (2,-3) -- (2.5,-3.5);
      \draw (2,-4) -- (2.5,-3.5);
      \draw (2,-4) -- (2.5,-4.5);
      \draw (3,-4) -- (2.5,-4.5);
      \draw (3,-4) -- (3.5,-4.5);
      \draw (4,-4) -- (3.5,-4.5);
      \draw (4,-4) -- (5,-5);
      
      \draw[fill] (0,0) circle [radius=0.03];
      \draw[fill] (0.5,-0.5) circle [radius=0.03];
      \draw[fill] (0,-1) circle [radius=0.03];
      \draw[fill] (0.5,-1.5) circle [radius=0.03];
      \draw[fill] (1,-1) circle [radius=0.03];
      \draw[fill] (1.5,-0.5) circle [radius=0.03];
      \draw[fill] (2,-1) circle [radius=0.03];
      \draw[fill] (1.5,-1.5) circle [radius=0.03];
      \draw[fill] (1,-2) circle [radius=0.03];
      \draw[fill] (1.5,-2.5) circle [radius=0.03];
      \draw[fill] (2,-2) circle [radius=0.03];
      \draw[fill] (2.5,-2.5) circle [radius=0.03];
      \draw[fill] (2,-3) circle [radius=0.03];
      \draw[fill] (2.5,-3.5) circle [radius=0.03];
      \draw[fill] (2,-4) circle [radius=0.03];
      \draw[fill] (2.5,-4.5) circle [radius=0.03];
      \draw[fill] (4,-4) circle [radius=0.03];
      \draw[fill] (4.5,-4.5) circle [radius=0.03];
      \draw[fill] (5,-5) circle [radius=0.03];
      \draw[fill] (3,-4) circle [radius=0.03];
      \draw[fill] (3.5,-4.5) circle [radius=0.03];
      \node [below] at (2.5,-5.5) {\parbox{0.3\linewidth}{\caption*{(b) Random reversals (RR)}\label{fig:RR1}}};
\end{scope}

      \draw[-{Latex[length=3mm, width=1mm]}]  (6,0) to node [midway,fill=white, text height=1ex] {Field} (6,-2);

\end{tikzpicture}
\caption{Models for studying transport in disordered media}
\label{fig:RC-RR}
\end{figure}

In this paper, we study fluctuations in time taken to cover a macroscopic distance in the RC model with quenched disorder (branch lengths do not change in time). We compute the variance of transit times and show that it diverges at a critical bias even in the RC implying that there is indeed a second regime where the variance of transit times is anomalously large although the drift velocity is finite. This is established by considering a generalized RC model with different strengths of bias along the backbone and the branches and analyzing the dynamics exactly. In the limit where only forward movement is allowed along the backbone, we compute two distinct types of variances in transit times: (a) $\sigma_{temporal}^2$, coming from history-to-history fluctuations and (b) $\sigma_{disorder}^2$, coming from sample-to-sample fluctuations of the history-averaged mean transit time. The steady state considerations that allowed $v_{drift}$ to be found in \citep{White_1984}, can be used to find $\sigma_{disorder}^2$ but not the variance $\sigma_{temporal}^2$, which is relevant for the practical case of a single realization of quenched disorder. To compute the latter, we need to follow the path of a single walker, along the RC, allowing for multiple excursions into each branch, as studied in \citep{POTTIER19951} and \citep{BALAKRISHNAN19951}, but circumventing the approximation made in these works. We find that $\sigma_{temporal}^2$ and $\sigma_{disorder}^2$ are different functions of the bias, yet they diverge at the same critical bias, indicating the onset of a regime where fluctuations are anomalously large.

\begin{figure}[!htb]
     \centering
     \begin{subfigure}{0.49\textwidth}
         \centering
         \includegraphics[width=\textwidth]{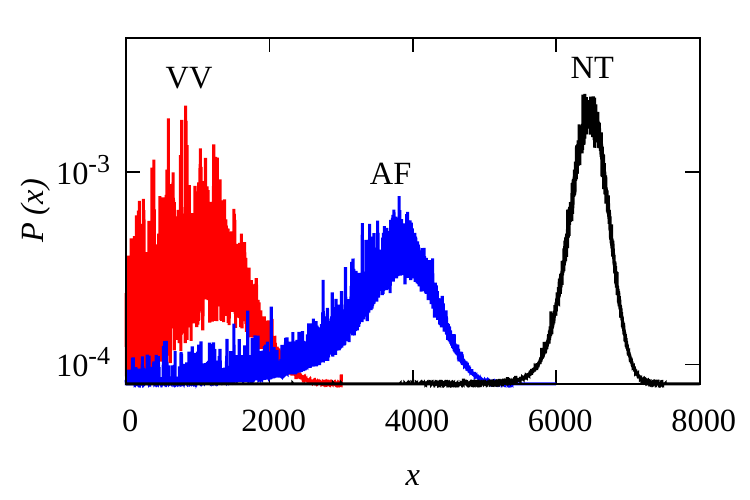}
         \caption{Particle position profile on backbone}
         \label{fig:xprof-1}
     \end{subfigure}
     \hfill
     \begin{subfigure}{0.49\textwidth}
         \centering
         \includegraphics[width=\textwidth]{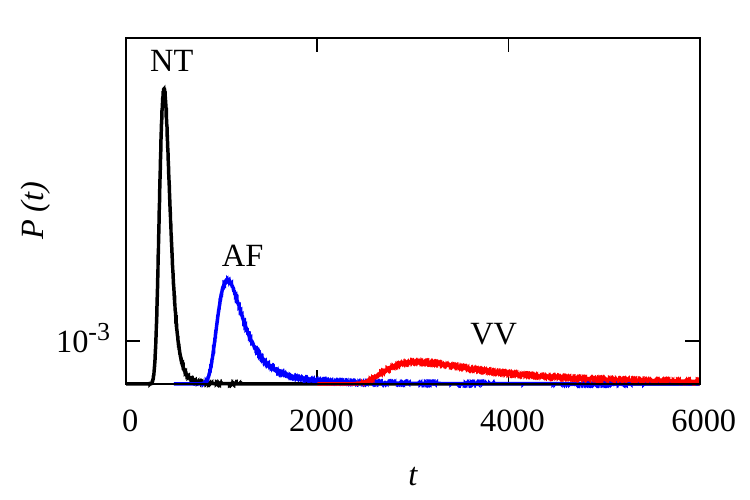}
         \caption{Transit time probability density}
         \label{fig:tprof-1}
     \end{subfigure}
        \caption{Profiles of particle position and transit time in the three regimes: (i) Normal Transport (black), (ii) Anomalous Fluctuation (blue) and (iii) Vanishing Velocity (red). (a) In the Normal Transport and the AF regimes the particle covers a significant distance at $t=6400 s$ while in the VV regime the displacement is negligible. (b) The transit time probability density at $N = 460$ follows normal behaviour in the Normal Transport regime, whereas in the AF and the VV regimes it has a fat tail. In (b) the distributions for the AF and VV regimes are shifted by $500$ and $2000$ units along the x-axis, respectively.}
        \label{fig:xtprof}
\end{figure}

Thus there are three regimes of transport through the RC: (a) the Normal Transport regime, for small values of bias where $v_{drift}$ and $\sigma_{temporal}^2$ are finite, (b) the Anomalous Fluctuation (AF) regime, which occurs for intermediate values of bias where $v_{drift}$ is non-zero but $\sigma_{temporal}^2$ diverges and (c) the Vanishing Velocity (VV) regime where $v_{drift}$ is zero and $\sigma_{temporal}^2$ remains divergent. Figure \ref{fig:xtprof} shows the differences in the profiles of particle position and distribution of transit times in these three regimes.

\begin{figure}[t]
    \centering
    \begin{tikzpicture}[scale=0.85]
      \draw[-{Latex[length=2mm, width=1.5mm]}, thick] (0,-0.2)--(0,5.3);
      \draw[-{Latex[length=2mm, width=1.5mm]}, thick] (-0.2,0)--(9.35,0);
      
      \fill[pattern=north east lines, pattern color=lightgray] (3,0) rectangle (6,5);
      \fill[pattern=crosshatch, pattern color=darkgray] (6,0) rectangle (9,5);
      
      \draw[dashed, ultra thick, black!60!green] (3,2.5) --(3,5);
      \draw[ultra thick, black!30!red] (3,0) --(3,2.5);
      \draw[dashed, ultra thick, black!60!green] (6,2.5)--(6,5);
      \draw[ultra thick, black!30!red] (6,0)--(6,2.5);
      \draw[dotted, thick] (9,0)--(9,5);
      \draw[dashed] (-0.1,2.5)--(9,2.5);
      \draw (3,-0.2)--(3,0);
      \draw (6,-0.2)--(6,0);
      \draw (9,-0.2)--(9,0);
      
      \node[below] at (0,-0.3) {$x\to\infty$};
      \node[below] at (0,-0.75) {$g = 0$};
      \node[below] at (3,-0.3) {$x=2$};
      \node[below] at (6,-0.3) {$x=1$};
      \node[below] at (9,-0.3) {$x=0$};
      \node[below] at (9,-0.75) {$g=1$};
      \node[right] at (9.3,0) {$x$};
      \node[left] at (-0.3,0) {$\lambda=0$};
      \node[left] at (-0.3,2.5) {$\lambda=1$};
      \node[above left] at (0,5) {$\lambda$};
      
      \node [xshift=-0.5cm,rotate=90,anchor=north] at (3,1.25) {{\footnotesize Second order}};
      \node [xshift=-0.5cm,rotate=90,anchor=north] at (3,3.75) {{\footnotesize First order}};
      \node [xshift=-0.5cm,rotate=90,anchor=north] at (6,1.25) {{\footnotesize Second order}};
      \node [xshift=-0.5cm,rotate=90,anchor=north] at (6,3.75) {{\footnotesize First order}};
      
      \node[align=center, above, rectangle, draw] at (1.5,5.1)%
      {\small Normal\\ \small Transport\\ \small $v_{drift}>0$ \\ \small $\sigma_{temp}^2<\infty$};
      \node[align=center, above, rectangle, draw] at (4.5,5.1)%
      {\small Anomalous\\ \small Fluctuation\\ \small $v_{drift}>0$ \\ \small $\sigma_{temp}^2\to\infty$};
      \node[align=center, above, rectangle, draw] at (7.5,5.1)%
      {\small Vanishing\\ \small Velocity\\ \small $v_{drift}=0$ \\ \small $\sigma_{temp}^2\to\infty$};
      
    \end{tikzpicture}
    \caption{Depiction of the different regimes of transport in the RC. Here $x = L_g/\xi$ with $L_g$ and $\xi$ being the lengths which characterise the bias and disorder realizations respectively. $\lambda$ is the power which appears in the disorder distribution Eq. (\ref{eq:Gammar}).}
    \label{fig:phasediag-lmu}
\end{figure}
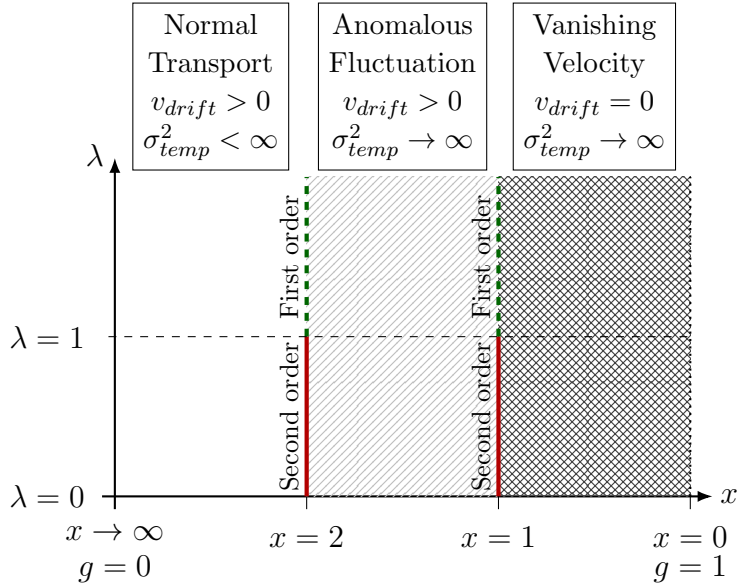

A second issue addressed in this paper concerns the order of the transition to the strong fluctuation regime. In \citep{Demaerel_2018}, it was pointed out that the transition of $v_{drift}$ to zero may be continuous or first order, depending on whether the distribution of branch lengths is exponential (as in \citep{MBDDhar1983,White_1984}), or whether there are strong enough power-law corrections to the exponential form characterised by the exponent $\lambda$ in Eq. (\ref{eq:Gammar}). Interestingly, the variance $\sigma_{temporal}^2$ shows a transition (first-order or continuous) from a finite value to infinity, with the order of the transition being the same as the transition to $v_{drift} = 0$, even though the two transitions occur at quite different values of the bias. Figure \ref{fig:phasediag-lmu} depicts the three regimes of transport along with the order of transitions to the AF and VV regimes, as a function of bias and exponent $\lambda$.

Finally, we also analyze the fluctuations in transit time on the bond diluted Bethe lattice. We analyze only $\sigma_{disorder}^2$ in this case and show that it diverges at a critical value of the bias, implying the existence of a second anomalous regime in this case as well.

\section{Random Comb (RC) Model}\label{sec:RC}
\subsection{Distribution of branch lengths}
A comb is a structure consisting of a one-dimensional backbone with branches. Combs with branches of infinite length are known to lead to anomalous transport properties for random walkers \citep{WEISS1986474}. The random comb (RC) under consideration here is a disordered structure with branches of random lengths much like the infinite cluster in percolation (Fig. \ref{fig:RC-RR}). A realization of disorder, $\textbf{R}$, is specified by the set of lengths $\{L_1,L_2,\dotsc,L_N\}$ of branches which are attached to a backbone of $N$ sites. These lengths are drawn independently from a distribution $\Gamma(L)$, implying that the probability of occurrence of the configuration is $P(\textbf{R}) = \prod_{i=1}^N\Gamma(L_i)$. The choice of $\Gamma(L)$ is motivated by the form of the two-point correlation function $C(r)$ in the percolation problem, thus taken to be

\begin{equation} \label{eq:Gammar}
    \Gamma(L) = \frac{1}{Z}\,\frac{e^{-L/\xi}}{1+(L/L_0)^\lambda}
\end{equation}
where $\xi$ is the correlation length and $Z = \sum_{L=0}^{\infty}e^{-L/\xi}[1+(L/L_0)^\lambda]^{-1}$ is a normalization constant.

The above form for $\Gamma(L)$, proposed earlier in \cite{Demaerel_2018}, is motivated by the Ornstein-Zernike (OZ) form for $C(r)$ for $p > p_c$. The OZ value of $\lambda$ is $(d-1)/2$ for spatial dimensions $d \geq 3$ and $\lambda = 2$ for $d = 2$ \cite{campanino2009finite,Campanino2011,Campanino2016}. On the RC it is instructive to consider the effect of different values of $\lambda$ as it has a significant effect on the nature of the transition as discussed in Section \ref{sec:OZcorr}. The case $\lambda = 0$, corresponding to a purely exponential decay, was the form assumed in \cite{MBDDhar1983,White_1984}, while the case $\lambda = 2$ was studied in \cite{Demaerel_2018}.

A quantity $A(\textbf{R},t)$ may have different values in every temporal evolution, or history, of particle motion in the same realization $\textbf{R}$ with the same initial conditions. The average over histories for a given $\textbf{R}$ is denoted by $\overline{A(\textbf{R},t)}$. Subsequently, one may take an average over the disorder configurations, defined as $\langle\overline{A(t)}\rangle = \sum_{\textbf{R}} \overline{A(\textbf{R},t)}P(\textbf{R})$.

\subsection{Particle Dynamics}\label{subsec:pardyn}

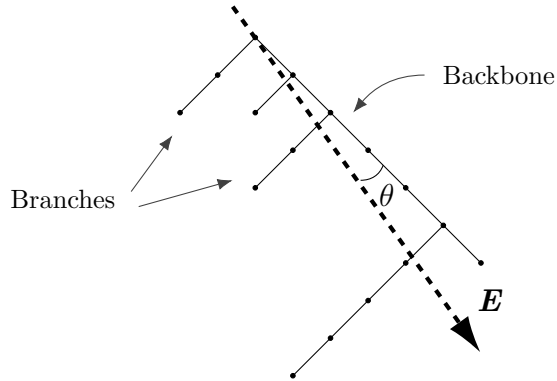
\begin{figure}[t]
    \centering
    \begin{tikzpicture}[scale=1]
      \draw (0,0) -- (3,-3);
      \draw[fill] (0,0) circle [radius=0.03];
      \draw[fill] (0.5,-0.5) circle [radius=0.03];
      \draw[fill] (1,-1) circle [radius=0.03];
      \draw[fill] (1.5,-1.5) circle [radius=0.03];
      \draw[fill] (2,-2) circle [radius=0.03];
      \draw[fill] (2.5,-2.5) circle [radius=0.03];
      \draw[fill] (3,-3) circle [radius=0.03];
      
      \draw (0,0) -- (-1,-1);
      \draw (0.5,-0.5) -- (-0.5+0.5,-0.5-0.5);
      \draw (1,-1) -- (-1+1,-1-1);
      \draw (2.5,-2.5) -- (-2+2.5,-2-2.5);
      
      \draw[fill] (-0.5,-0.5) circle [radius=0.03];
      \draw[fill] (-1,-1) circle [radius=0.03];
      \draw[fill] (0,-1) circle [radius=0.03];
      \draw[fill] (0.5,-1.5) circle [radius=0.03];
      \draw[fill] (0,-2) circle [radius=0.03];
      \draw[fill] (2,-3) circle [radius=0.03];
      \draw[fill] (1.5,-3.5) circle [radius=0.03];
      \draw[fill] (1,-4) circle [radius=0.03];
      \draw[fill] (0.5,-4.5) circle [radius=0.03];
      
      \draw [dashed, ultra thick,-{Latex[length=5mm, width=3mm]}] (-0.3,0.41) -- (3,-4.2);
      \draw (1.7,-1.7) to [out=250,in=0] (1.40,-1.9);
      
      \node[below right] at (1.5,-1.8) {$\theta$};
      \node[right] at (1.75+0.6,-0.5) {\footnotesize Backbone};
      \draw[-{Latex[length=2mm, width=1.5mm]}, darkgray] (1.75+0.5,-0.5) to [out=180,in=50] (1.2+0.1,-1.1+0.1);
      
      \node[right] at (-3-0.4,-1.65-0.5) {\footnotesize Branches};
      \draw[-{Latex[length=2mm, width=1.5mm]}, darkgray] (-0.8-0.25-0.6,-1.65-0.4) -- (-1-0.1,-0.8-0.2-0.3);
      \draw[-{Latex[length=2mm, width=1.5mm]}, darkgray] (-0.8-0.25-0.5,-1.75-0.5) -- (-0.05-0.25,-1.9);
      
      \node[right] at (2.8,-3.5) {\textbf{\textit{E}}};
      
    \end{tikzpicture}
    \caption{Generalized RC model with different components of bias along the backbone and the branches.}
    \label{fig:RCField}
\end{figure}

We model the motion of particles on the disordered medium by a random walk. In the absence of a driving field, the particles undergo diffusion on the lattice \cite{ASLANGUL1994533}. The effect of an applied field is modelled by making the random walk biased, with a larger transition rate along the field than against it. The value of the bias in the backbone and branches may differ, reflecting the fact that the applied field may have different components along the two (Fig. \ref{fig:RCField}).

\begin{figure}[t]
    \centering
    \begin{tikzpicture}
    \draw (0,0)--(9,0);
    \draw (0,0)--(0,-1.5);
    \draw (1.5,0)--(1.5,-3);
    \draw (7.5-1.5,0)--(7.5-1.5,-1.5);
    \draw (6-1.5,0)--(6-1.5,-6);
    \draw (10.5-1.5,0)--(10.5-1.5,-3);
    \draw[fill] (0,0) circle [radius = 0.05];
    \draw[fill] (0,-1.5) circle [radius = 0.05];
    \draw[fill] (1.5,0) circle [radius = 0.05];
    \draw[fill] (1.5,-1.5) circle [radius = 0.05];
    \draw[fill] (1.5,-3) circle [radius = 0.05];
    \draw[fill] (3,0) circle [radius = 0.05];
    \draw[fill] (7.5-1.5,-1.5) circle [radius = 0.05];
    \draw[fill] (6-1.5,0) circle [radius = 0.05];
    \draw[fill] (6.0-1.5,-1.5) circle [radius = 0.05];
    \draw[fill] (6.0-1.5,-3) circle [radius = 0.05];
    \draw[fill] (6.0-1.5,-4.5) circle [radius = 0.05];
    \draw[fill] (6.0-1.5,-6.0) circle [radius = 0.05];
    \draw[fill] (7.5-1.5,0) circle [radius = 0.05];
    \draw[fill] (9-1.5,0) circle [radius = 0.05];
    \draw[fill] (10.5-1.5,0) circle [radius = 0.05];
    \draw[fill] (10.5-1.5,-1.5) circle [radius = 0.05];
    \draw[fill] (10.5-1.5,-3) circle [radius = 0.05];
    
    
    
    \node[above] at (0,0.5) {$(1,0)$};
    \node[above] at (1.5,0.5) {$(2,0)$};
    \node[above] at (3,0.5) {$\dotsc$};
    \node[above] at (6-1.5,0.5) {$(n,0)$};
    \node[above] at (9-1.5,0.5) {$\dotsc$};
    \node[above] at (10.5-1.5,0.5) {$(N,0)$};
    \node[left] at (5.8-1.5,-4.5) {$(n,m)$};

    \draw[-{Latex[length=3mm, width=1mm]}] (6.3-1.5,-0.2-1.5)--(6.3-1.5,-1.3-1.5);
    \draw[-{Latex[length=3mm, width=1mm]}] (5.7-1.5,-1.0-1.5)--(5.7-1.5,-0.5-1.5);
    \draw[-{Latex[length=3mm, width=1mm]}] (6.3-1.5,-0.2-1.5-3)--(6.3-1.5,-1.3-1.5-3);
    \draw[-{Latex[length=3mm, width=1mm]}] (5.7-1.5,-1.0-1.5-3)--(5.7-1.5,-0.5-1.5-3);
    \draw[-{Latex[length=3mm, width=1mm]}] (6.2-1.5,0.2)--(7.3-1.5,0.2);
    \draw[-{Latex[length=3mm, width=1mm]}] (6.2-1.5-1.5,0.2)--(7.3-1.5-1.5,0.2);
    \node[right] at (6.3-1.5,-0.75-1.5) {$u$};
    \node[left] at (5.7-1.5,-0.75-1.5) {$v$};
    \node[right] at (6.3-1.5,-0.75-1.5-3) {$u$};
    \node[left] at (5.7-1.5,-0.75-1.5-3) {$v$};
    \node[above] at (6.75-1.5,0.2) {$w$};
    \node[above] at (8.25-3-1.5,0.2) {$w$};
    
    
    \node[right] at (6-1.5,-3) {\footnotesize{$\gamma = u+v$}};
    \node[below] at (6-1.5,-6) {\footnotesize{$\gamma = v$}};
    \node[above] at (1.5,0) {\footnotesize{$\gamma = u+w$}};
    \node[above] at (7.5,0) {{\footnotesize$\gamma = w$}};

    \end{tikzpicture}
    \caption{In the forward transport model ($w' = 0$) the transition rates, $u$ and $v$ along branch bonds and $w$ along the backbone bonds, are uniform. This corresponds to the inverse of mean waiting time $\gamma$ of a random walker being different on different sites.}
    \label{fig:RCrates}
\end{figure}
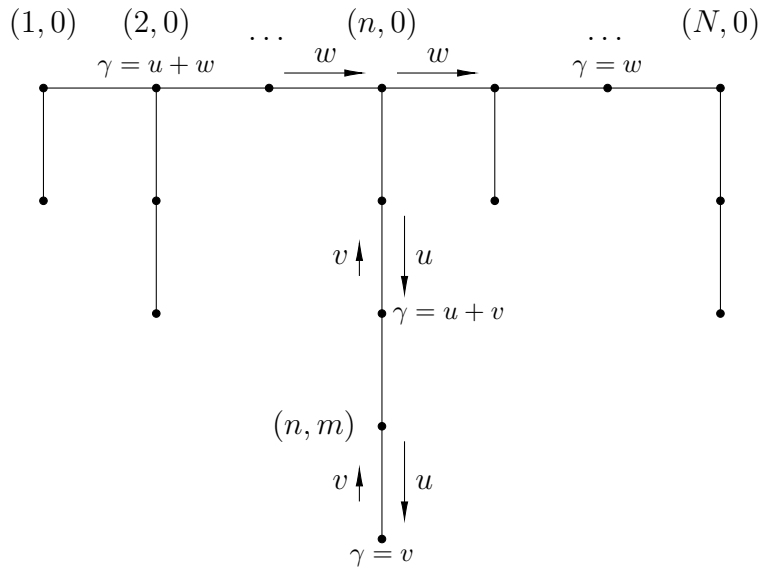

Let $n = 1, 2, ..., N$ be the coordinate of a site along the backbone, to which is attached a branch of length $L_n$, and let $m = 0, 1, ..., L_n$ label the sites in the branch, with $m = 0$ corresponding to the backbone site (Fig. \ref{fig:RCrates}). Let the rate of transition between any two sites $a$ and $b$ be $W_{a,b}$. Therefore, $W_{n,n+1} = w$ and $W_{n,n-1} = w'$ are the forward and the backward transition rates along the backbone, and $W_{m,m+1} = u$ and $W_{m,m-1} = v$ are the forward and the backward rates, in the branches, respectively. Let $p_{n,m}(t)$ be the probability that a particle is on site $(n,m)$ of the RC at time $t$. If $L_n$ is non-zero, the time evolution of $p_{n,m}(t)$ follows the equations
\begin{equation} \label{eq:MastEqn}
\begin{split}
    \frac{dp_{n,0}(t)}{dt} &= w\,p_{n-1,0}(t) + w'\,p_{n+1,0}(t) + v\,p_{n,1}(t) - (w + u + w')\,p_{n,0}(t), \\
    \frac{dp_{n,m}(t)}{dt} &= u\,p_{n,m-1}(t) + v\,p_{n,m+1}(t) -(u+v)p_{n,m}(t),\quad 0<m<L_n,\\
    \frac{dp_{n,L_n}(t)}{dt} &= u\,p_{n,L_n - 1}(t) - v\,p_{n,L_n}(t).
\end{split}
\end{equation}
If $L_n = 0$, the evolution follows
\begin{equation}\label{eq:MastEqn2}
    \frac{dp_{n,0}(t)}{dt} = w\,p_{n-1,0}(t) + w'\,p_{n+1,0}(t) - (w+w')\,p_{n,0}(t).
\end{equation}

The rates of hopping ($u,v,w$ and $w'$) are related to the strength of the field $E$ and temperature $T$ through:
\begin{equation}\label{eq:ratesET}
\begin{split}
    \frac{v}{u} &= e^{-Ea\sin\theta/k_B T}, \\
    \frac{w'}{w} &= e^{-Ea\cos\theta/k_B T}
\end{split}
\end{equation}
where $a$ is the lattice spacing (taken to be unity in the rest of the paper). The bias in a branch may be parameterized by $g$, defined through $u = W_0(1+g)$, $v = W_0(1-g)$, where $W_0$ is a constant. We assume a uniform set of transition rates $(u,v,w$ and $w')$ which ensures that the particle density is uniform in the backbone in the steady state.

\subsection{Forward Transport Model}\label{subsec:FTM}
The general problem of transport on the RC involves calculating the total transit time along the backbone. Now any revisit to a site on the backbone (after moving away from it, along the backbone) entails remembering the length of the branch on the site in question, which makes the process non-Markovian. In the spirit of a commonly used approximation, namely the `continuous-time random walk' (CTRW) as used in \citep{MontrollWeiss1965,ScherLax1973}, the non-Markovian property may be avoided by considering a process in which a fresh branch length is drawn from the given distribution, at each revisit to the site \citep{BALAKRISHNAN19951,POTTIER19951}. One may ask how good the CTRW approximation is. For a certain problem with disordered hopping rates, it was shown that this approximation gives the exact answer for the \textit{average probability} of occupying a site at a certain time, but cannot properly account for the variance of the probability \citep{KlafterSilbey1980}. A similar conclusion is likely to hold for the CTRW-like approximation for the random comb.

Our approach is different; we consider a \textit{forward transport limit} in which the problem of memory does not arise at all: the particle is allowed to hop only in the forward direction along the backbone so that revisits to a backbone site $n$ are not allowed once the particle has jumped to site $n + 1$. This allows exact results to be obtained and obviates the necessity of making a CTRW-like approximation in which a fresh length of the branch is drawn on every revisit to site $n$. We work with the FTM as it offers simplicity and tractability, avoiding any approximation, yet keeps the essence of the problem, namely the effects of quenched disorder on dynamics, intact.

The FTM may be realized as follows. Consider applying a field (Fig. \ref{fig:RCField}), aligned almost along the backbone so that $\theta$ is very small. Now consider the joint limit $\theta\to 0$, $E\to\infty$, such that $E\theta\sim$ constant. This leads to $w'/w \simeq 0$, hence disallowing backward motion along the backbone. (Another way of achieving this limit is to keep the field constant and reduce the temperature of the system $T\to 0$ with $\theta\to 0$ such that $\theta/T \sim$ constant.) Examples of systems where a ``no-return" condition may be met are hydrodynamic dispersion in porous media \cite{BOUCHAUD1990127} and high-performance liquid chromatography \cite{DeVos2016}.

In the FTM, the hopping rate $w'$ in Eqs. (\ref{eq:MastEqn}) and (\ref{eq:MastEqn2}) vanishes. The rates $u, v$ and $w$ may be related to the dynamics of a single particle on the RC as illustrated in Fig. \ref{fig:RCrates}. Let $\gamma$ be the rate for particle movement from a particular site ($\gamma$ is site-dependent). For any branch site except the last, $\gamma =u+v$, with a probability $p=u/(u+v)$ for a downward and probability $q = v/(u+v)$ for an upward hop. At the last site in any branch, $\gamma =v$, and the probability of an upward hop is $1$. Likewise, on the backbone, for a site with no branch, $\gamma=w$ with probability $1$ for rightward motion, whereas on a site with a branch, $\gamma =u+w$ with a probability $r=u/(u+w)$ of entering the branch, and $(1-r)=w/(u+w)$ of moving rightward along the backbone. This choice of inhomogeneous values of $\gamma$ for the walker corresponds to uniform transition rates on backbone bonds - the feature which ensures uniform density along the backbone, allowing the current to be calculated in steady state. By contrast, if $\gamma$ is chosen to be uniform (as in \cite{BALAKRISHNAN19951}), the transition rates on backbone bonds are non-uniform, and determining the density profile is a nontrivial problem. The distribution $\phi(t)$ of the hopping times is given by
\begin{equation} \label{eq:phi}
    \phi(t) = \gamma e^{-\gamma t}.
\end{equation}
\section{Fluctuations of the transit time}\label{sec:analytical-1}
While characterizing transport, we may ask either for the distribution of the location of a particle at a specified time $t$, or for the distribution of the time taken for a particle to reach a specified site. In general, it is not straightforward to go between these two descriptions \citep{CaceresPRA41}, and we choose to work with the latter.

In this section, we first discuss the steady state density profile on the RC, and how it may be used to find $v_{drift}$. However, this approach does not suffice to compute the variance $\sigma_{temporal}^2$ (defined in Sec. \ref{sec:def}) of transit times, which is the central quantity of interest in this paper. We compute it in the FTM by following the trajectory of a walker along the backbone by including multiple excursions into branches.

\subsection{Steady state: Calculation of $v_{drift}$}
Consider the forward transport model on a random comb with a transition rate $w$ between sites $(n,n+1)$ of the backbone $n = 1,2,\dotsc,N_s$, with periodic boundary conditions. In the steady state reached at long times, the current vanishes in every branch due to reflecting boundaries at branch tips, whereas it has a uniform value $J$ in every bond of the backbone. This suffices to find the full density profile, which in turn determines the transit time, $T_{N_s}(\textbf{R})$, taken by a particle to traverse the $N_s$ sites of the backbone in a given realization $\textbf{R}$. Following \cite{White_1984}, if $\rho_0$ is the density at site $(n,0)$, we have $\rho_{0} = J/w$ for all $n$. The zero-current condition in the branch implies that, $\rho_{m} = \rho_0\,e^{m/L_{g}}$, where
\begin{equation}\label{eq:40}
    L_{g} = \frac{1}{\log{(u/v)}}
\end{equation}
is a bias-induced length scale which describes the exponential growth of density in the branches. The mean transit time in the steady state is related to the overall number of particles $\mathcal{N}$ by $\overline{T}_{N_s}(\textbf{R}) = \mathcal{N}/J$. Thus as in \cite{White_1984}, we find
\begin{equation}\label{eq:50}
    \overline{T}_{N_s}(\textbf{R}) = \sum_{i=1}^{N_s}\overline{t}_i
\end{equation}
where,
\begin{equation}\label{eq:52}
    \overline{t}_i =  \frac{1}{w}\frac{\exp[(L_i+1)/L_g]-1}{\exp(1/L_g)-1}.
\end{equation}
So long as the disorder average of $\overline{t}_i$, over the distribution (\ref{eq:Gammar}), is finite (i.e. if $L_g>\xi$), the law of large numbers holds and implies that $T_{N_s}(\textbf{R})/N_s$ converges to $\langle T_{N_s}(\textbf{R})\rangle/N_s$ as $N_s\to\infty$. The answer is independent of $\textbf{R}$ and for the case of an exponentially decaying distribution of branch lengths (i.e. $\lambda = 0$ in (\ref{eq:Gammar})) is given by
\begin{equation}\label{eq:41}
\begin{aligned}
    \frac{\langle\overline{T}_{N_s}\rangle}{N_s} &= \frac{1}{w\big[ 1 - \exp{(1/L_{g}-1/\xi)} \big]},  &L_{g} > \xi,\\
    &= \infty, &L_{g} \leq \xi.
\end{aligned}
\end{equation}

Correspondingly, the mean drift velocity is given by $v_{drift} = N_s/\langle \overline{T}_{N_s}\rangle$. Therefore, in the steady state, in the forward transport limit for the random comb, we have
\begin{equation}\label{eq:42}
\begin{aligned}
    v_{drift} &= w\big[ 1-\exp(1/L_{g}-1/\xi) \big], &L_{g}>\xi,\\
    &= 0, &L_{g}\leq\xi.
\end{aligned}
\end{equation}
The condition $v_{drift} = 0$ defines the VV regime.

\subsection{Fluctuations of transit times: Calculation of $\sigma_{temporal}^2$}\label{sec:def}
The variances $\sigma_{temporal}^2$ and $\sigma_{disorder}^2$ are defined as follows
\begin{equation}\label{eq:stempdef}
\begin{split}
    \sigma_{temporal}^2 &= \langle \overline{T_N^2(\textbf{R})} \rangle - \langle \overline{T_N(\textbf{R})}^2 \rangle \\
    \sigma_{disorder}^2 &= \langle \overline{T_N(\textbf{R})}^2\rangle - \langle \overline{T_N(\textbf{R})}\rangle^2.
\end{split}
\end{equation}

The steady state approach suffices to calculate $\sigma_{disorder}^2$ (the variance due to sample-to-sample fluctuations of the mean transit time across realizations of disorder) but not $\sigma_{temporal}^2$ (the variance due to history-to-history fluctuations of the total transit time in realization \textbf{R}, finally averaged over all \textbf{R}). To find this, we need to track the full trajectory of the particle through the system. We use the FTM, the key simplifying feature of which is that the total transit time $T_N(\textbf{R})$ is just the sum of residence times $\tau(L_n)$ on the individual backbone sites $n$, where $\tau(L_n)$ is the time interval between reaching site $n$ of the backbone and hopping from $n$ to $n+1$. During the time it is resident on site $n$, the particle may make multiple excursions into the attached branch if any. The residence time, therefore, involves the sum of the first-return times corresponding to each excursion. Therefore, we need the distribution of the first-return times on a single branch.

We assume the backbone to be infinite and follow the particle till it reaches the $N^{th}$ backbone site, with the initial condition that it is on the backbone site $n = 1$ at time $t = 0$. The variance of transit time in disorder realization $\textbf{R}$ due to history-to-history fluctuations is
\begin{equation}\label{eq:31}
    \sigma_{N}^2(\textbf{R}) = \overline{T_N^2(\textbf{R})} - \overline{T_N(\textbf{R})}^2.
\end{equation}
where $T_N(\textbf{R}) = \sum_{i=1}^N\tau(L_i)$. Let $\psi_L(t)$ denote the distribution of the residence times. A typical excursion into the branch involves the particle going from backbone site $m=0$ to  branch site $m=1$ after waiting for some time at $m = 0$. It may go further into the branch, finally returning to $m = 0$, marking the completion of one excursion, and may perform several such excursions at $n$ before hopping to $n+1$. Let $t^{(j)}(L)$ denote the total time taken for $j$ excursions in a branch of length $L$ and let $f^{(j)}_L(t)$ be the distribution of these times. Thus, $t^{(1)}(L)$ denotes the first return time to the backbone and $f^{(1)}_L(t)$ denotes the distribution of first return times. Let us, for clarity, denote them by $t_{fr}(L)$ and $f_L(t)$, respectively. The particle may finally wait on the backbone site after the $j$ excursions, before hopping to the next backbone site, therefore,
\begin{equation}\label{eq:47}
    \psi_L(t) = \sum_{j=0}^{\infty}\int_0^t dt_1\, r^j f^{(j)}_L(t_1)(1-r)\phi(t-t_1).
\end{equation}
where $r$ is the probability of making an excursion into the branch. Decomposing the total time taken for $j$ excursions into individual first return times, we have
\begin{equation}\label{eq:46}
    f^{(j)}_L(t) = \int_0^t dt_j\dotsc\int_0^{t_3}dt_2\int_0^{t_2}dt_1\, f_L(t_1)f_L(t_2 - t_1)\dotsc f_L(t-t_{j-1}).
\end{equation}

\subsection{First return times on a branch}\label{subsec:FRT}
Consider a single branch of length $L$ ($>0$). At $t = 0$, let a particle be at $m = 0$. The distribution of first return times, $f_L(t)$, is found in terms of $\phi(t)$, the waiting time density at $m = 0$, and $Q_L(t)$, the distribution of first passage times from branch site $(n,1)$ to the backbone site $(n,0)$:
\begin{equation}\label{eq:3}
    f_L(t) = \int_0^t dt_1\,\phi(t_1)Q_L(t-t_1).
\end{equation}
Let $\tilde{f}_L(s), \tilde{\phi}_L(s)$ and $\tilde{Q}_L(s)$ be the Laplace transforms of $f_{L}(t)$, $\phi(t)$ and $Q_L(t)$, respectively, for example
\begin{equation}
    \tilde{f}_{L}(s) = \int_0^{\infty}dt\,e^{-st}f_{L}(t).
\end{equation}
The convolution form of (\ref{eq:3}) implies that
\begin{equation}\label{eq:4}
    \tilde{f}_L(s) = \tilde{\phi}(s)\tilde{Q}_L(s)
\end{equation}
where
\begin{equation}\label{eq:phis}
    \tilde{\phi}(s) = \frac{\gamma}{\gamma + s}
\end{equation}
follows from (\ref{eq:phi}) and $\gamma = u + w$ at $m = 0$. The expression for $\tilde{Q}_L(s)$ was found exactly in \cite{Khantha1983},
\begin{equation}\label{eq:5}
    \tilde{Q}_L(s) = \Big(\frac{v}{u}\Big)^{1/2}\bigg[\frac{\sqrt{v/u}\sinh{L\zeta} - \sinh{(L-1)\zeta}}{\sqrt{v/u}\sinh{(L+1)\zeta} - \sinh{L\zeta}}\bigg]
\end{equation}
where
\begin{equation}\label{eq:6}
    \cosh{\zeta} = \frac{(u+v)}{\sqrt{4uv}}\Big(1 + \frac{s}{u+v}\Big).
\end{equation}
This result was later used in \citep{BALAKRISHNAN19951} within a CTRW-like approximation in which branch lengths are drawn freshly at each revisit to a site.

Let $\overline{t}_{fr}(L) = -\big[\partial\log\tilde{f}_L(s)/\partial s\big]_{s=0}$ be the history-averaged mean first return time. After simplification we obtain
\begin{equation}\label{eq:9}
    \overline{t}_{fr}(L) = \frac{(u/v)^L-1}{u-v} + \frac{1}{u+w}.
\end{equation}
Similarly, we find the variance of the first return times by calculating the second derivative of $\log\tilde{f}_L(s)$ at $s = 0$, $\sigma^2_{fr}(L) = \big[\partial^2\log\tilde{f}_L(s)/\partial s^2\big]_{s=0}$, leading to
\begin{equation}\label{eq:10}
    \sigma_{fr}^2(L) = \frac{(u+v)(u/v)^{2L} - 4L(u-v)(u/v)^{L} - (u+v)}{(u-v)^3} + \frac{1}{(u+w)^2}.
\end{equation}

\begin{figure}[!t]
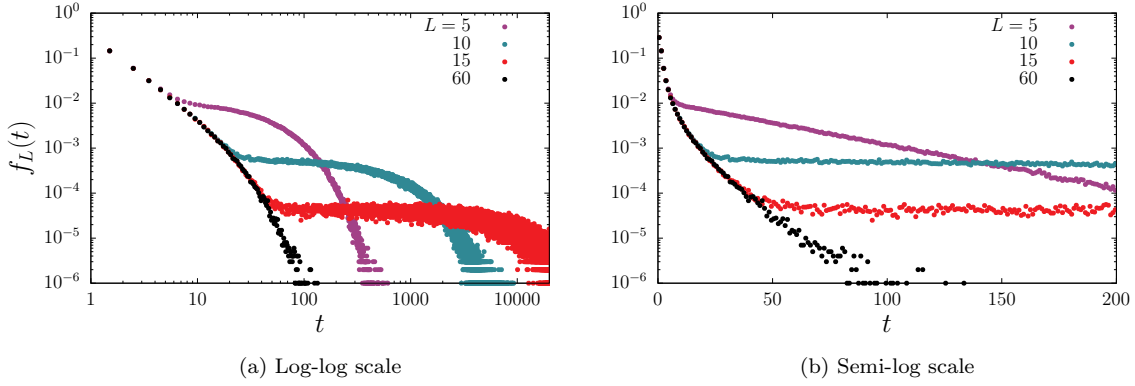

     \centering
     \begin{subfigure}{0.45\textwidth}
         \centering
         \scalebox{0.6}{\input{frt1.tex}}
         \caption{Log-log scale}
     \end{subfigure}
     \hfill
     \begin{subfigure}{0.45\textwidth}
         \centering
         \scalebox{0.6}{\input{frt2.tex}}
         \caption{Semi-log scale}
     \end{subfigure}
        \caption{Distribution of first return times for branches of different lengths for the same value of bias ($g=0.244$). The leftmost curve (black) is the distribution for an effectively infinite system and has a power-law (diffusive) and exponentially decaying (drift) regimes. Since a branch is finite, there is an additional ultra-slow exponentially decaying (trapping) regime. For times beyond $t=2\times 10^4$ the procedure discussed in \ref{sec:appendixA} is used.}
        \label{fig:frt}
\end{figure}

In the limit $L\to\infty$, the mean $\overline{t}_{fr}(L)$ and the variance $\sigma_{fr}^2(L)$ diverge, as expected for a biased random walk in an infinite 1D system. Interesting effects coming from the finiteness of the branch are seen in Monte Carlo simulations (Fig. \ref{fig:frt}). There are two time scales. The first scale $\sim (u+v)/2(u-v)$ corresponds to the crossover from diffusion to drift as the predominant mode of transport: the return times are distributed as a power-law $\sim t^{-3/2}$ in the diffusive regime whereas the distribution decays exponentially in the drift regime. There is a third trapping regime in which the return times follow $\sim \exp(-t/\tau_L)$, with $\tau_L \sim \exp(L/L_g)$. This is an outcome of the activation process required to clear the barrier against the `easy-direction' of bias. The very large times required to cross this barrier pose a challenge for numerical simulations, and a non-standard procedure had to be devised, as discussed in \ref{sec:appendixA}.

\subsection{Residence time on a backbone site}\label{subsec:ResT}
Applying the convolution theorem to (\ref{eq:47}), we get
\begin{equation}\label{eq:45}
    \tilde{\psi}_L(s) = \sum_{j=0}^{\infty}r^j(1-r)\tilde{f}^{(j)}_L(s)\tilde{\phi}(s).
\end{equation}
With $\tilde{f}^{(j)}_L(s) = \big[\tilde{f}_L(s)\big]^j$ from (\ref{eq:46}) and using (\ref{eq:4}), we obtain
\begin{equation}\label{eq:12}
    \tilde{\psi}_L(s) = \frac{(1-r)\tilde{\phi}(s)}{1-r\tilde{\phi}(s)\tilde{Q}_L(s)}.
\end{equation}
We find the mean and variance of the residence time by calculating the first and second derivatives of $\log \tilde{\psi}_L(s)$ at $s = 0$, with the results
\begin{equation}\label{eq:13}
    \overline{\tau}(L) = \frac{1}{w}\bigg[ \frac{(u/v)^{L+1}-1}{(u/v)-1} \bigg],
\end{equation}
\begin{multline}\label{eq:14}
    \sigma^2_{Res}(L) = \big[u^2(u-v+2w)(u/v)^{2L} - 2u(u-v)((2L+1)w+v)(u/v)^{L}\\ - v(v^2+2uw-uv)\big](u-v)^{-3} w^{-2}.
\end{multline}
Observing that the variance $\sigma_{Res}^2$ has contributions from $\sigma_{fr}^2$ as well as from the fluctuations due to different means corresponding to different $j$'s, we may write
\begin{equation}
    \sigma_{Res}^2(L) = \frac{1}{(u+w)^2} + \frac{r}{1-r}\sigma_{fr}^2(L) + \frac{r}{(1-r)^2}\overline{t}^2_{fr}(L).
\end{equation}
To highlight the $L$ dependence we may rewrite (\ref{eq:14}) as
\begin{equation}\label{eq:15}
    \sigma^2_{Res}(L) = Ae^{2L/L_{g}} - B((2L+1)w+v)e^{L/L_{g}} - C
\end{equation}
where $A = u^2(u-v+2w)[(u-v)^3 w^2]^{-1}$, $B = 2u[(u-v)^2 w^2]^{-1}$ and $C = v(v^2+2uw-uv)[(u-v)^3 w^2]^{-1}$.

\subsection{Total transit time across the backbone}
The history-averaged total transit time for a realization $\textbf{R}$ is $\overline{T}_N(\textbf{R}) = \sum_{i=1}^N\overline{\tau}(L_i)$. Using (\ref{eq:13}) we obtain the transit time, and note that it coincides with the steady state transit times (\ref{eq:50}) and (\ref{eq:52}). The variance of total transit time for a given realization is given by $\sigma_N^2(\textbf{R}) = \overline{T_N^2(\textbf{R})} - \overline{T_N(\textbf{R})}^2$, which implies
\begin{equation}
    \sigma_N^2(\textbf{R}) = \sum_{i=1}^N\sigma_{Res}^2(L_i).    
\end{equation}
As long as the average of $\sigma_{Res}^2(L)$ over the disorder distribution (\ref{eq:Gammar}) is finite, the law of large number holds and implies that $\sigma_N^2(\textbf{R})/N$ converges to $\langle\sigma_N^2(\textbf{R})\rangle/N$ as $N\to\infty$.

\subsection{Disorder averaging: Exponential distribution, $\lambda = 0$}
Consider the disorder distribution $\Gamma(L) = (1-e^{-1/\xi})e^{-L/\xi}$, corresponding to $\lambda = 0$ in (\ref{eq:Gammar}). The average of mean transit time over the disorder is given by $\langle \overline{T}_N\rangle =\sum_{\textbf{R}}$ $P(\textbf{R})\overline{T}_N(\textbf{R}) $ $= N\langle\overline{\tau}\rangle$, if $L_g>\xi$, where $\langle\overline{\tau}\rangle = \sum_{L}\overline{\tau}(L)\Gamma(L).$ Averaging (\ref{eq:13}), we reproduce the steady state result (\ref{eq:41}). It then follows that the mean drift velocity $N/\langle\overline{T}_N\rangle$ is identical to $v_{drift}$ in (\ref{eq:42}) from the steady state argument.

If $\langle \sigma_{Res}^2\rangle$ is finite, then using (\ref{eq:31}) and the fact that branches are independent and identically distributed, we have
\begin{equation}\label{eq:32}
    \frac{\sigma_{temporal}^2}{N} = \langle\sigma_{Res}^2\rangle.
\end{equation}
Using (\ref{eq:15}), we find
\begin{equation}\label{eq:43}
\begin{aligned}
    \frac{\sigma_{temporal}^2}{N} &= \frac{ 1 + (2w/v)e^{1/L_{g}-1/\xi} + (2w/v - 1)e^{2/L_{g}-2/\xi} }{w^2\big[1-e^{1/L_{g}-1/\xi}\big]^2\big[1-e^{2/L_{g}-1/\xi}\big]}, &L_{g}>2\xi,\\
    &= \infty, &L_{g}\leq 2\xi.
\end{aligned}
\end{equation}

\begin{figure}[!htb]
    \centering
    \includegraphics[width=0.45\textwidth]{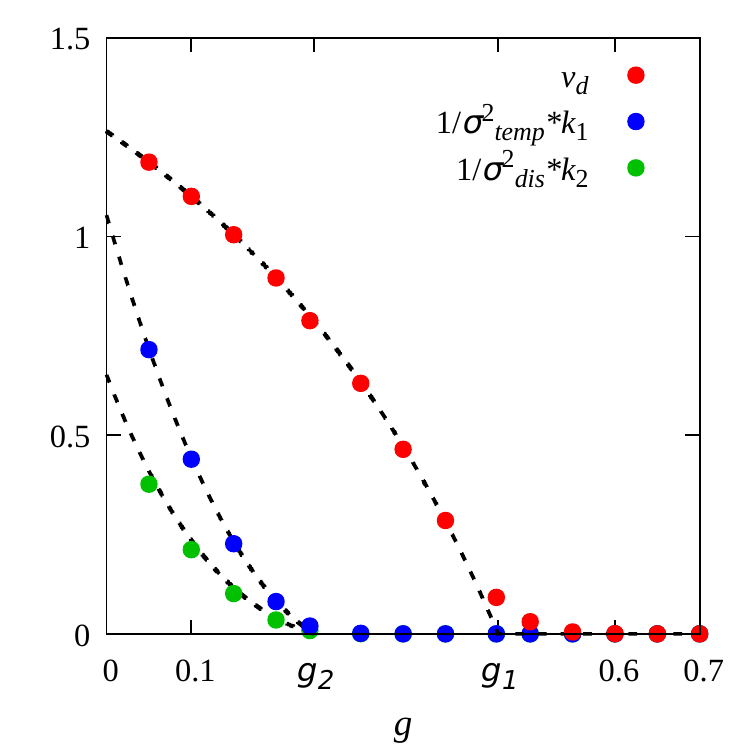}
    \caption{Mean drift velocity and variances $\sigma_{temporal}^2$ and $\sigma_{disorder}^2$ vs bias $g$. The critical values $g_1$ and $g_2$ mark the onset of the VV and AF regimes respectively. The dots represent the values from Monte Carlo simulations and the dashed black lines represent the theoretically expected curves (Eqs. (\ref{eq:42}), (\ref{eq:43}) and (\ref{eq:44})) with $N = 2000$. The two curves on the left are scaled by factors $k_1 = 3N$ and $k_2 = 3N/20$ for clarity.}
    \label{fig:vg-sec3}
\end{figure}

\begin{figure}[!htb]
     \centering
     \begin{subfigure}{0.45\textwidth}
         \centering
         \includegraphics[width=\textwidth]{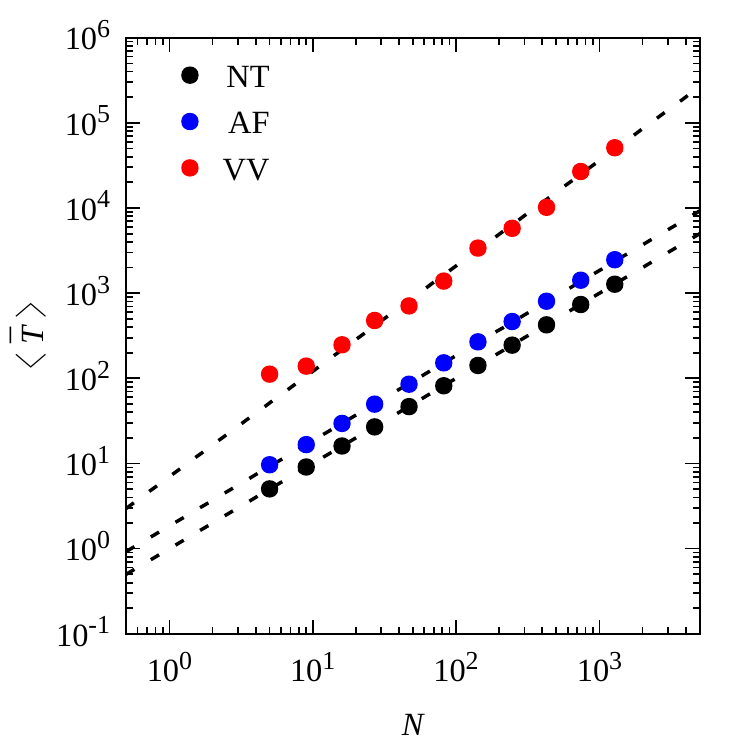}
         \caption{Mean transit time $\langle \overline{T}_N\rangle$ vs $N$}
         \label{fig:LoLN}
     \end{subfigure}
     \hfill
     \begin{subfigure}{0.45\textwidth}
         \centering
         \includegraphics[width=\textwidth]{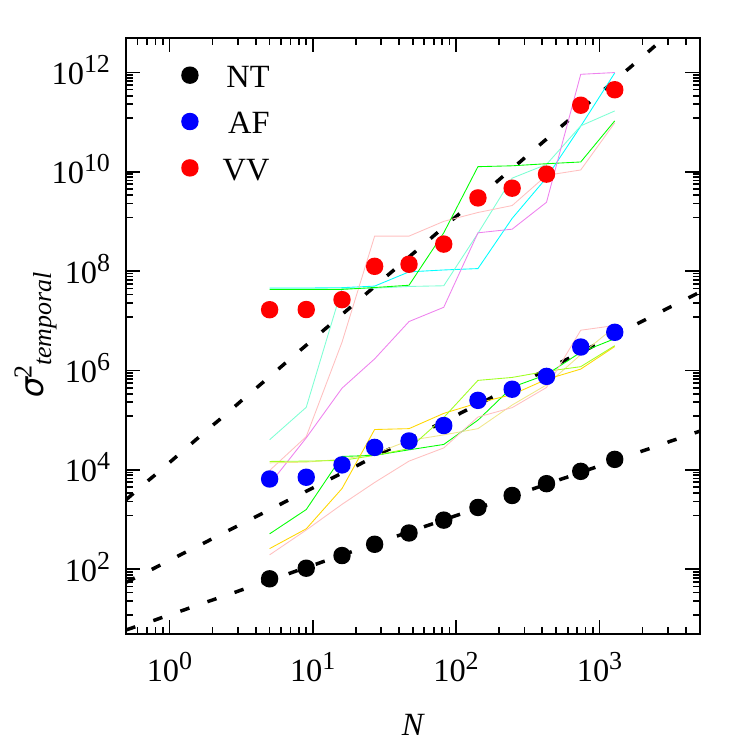}
         \caption{Variance $\sigma_{temporal}^2$ vs $N$}
         \label{fig:CLT}
     \end{subfigure}
        \caption{Breakdown of the central limit theorem in AF regime and breakdown of the law of large numbers in VV regime. The solid dots represent the values from Monte Carlo simulations averaged over $10^3$ realizations and $6.4\times10^5$ histories. (a) The dashed lines represent $y \sim N$ in the Normal Transport and the AF regimes, and $y \sim N^x$ in the VV regime. (b) The dashed lines represents $N$ in the Normal Transport and $N^{2x}$ in the AF and the VV regimes. In (b) each of the solid lines represent disorder average of $\sigma_{temporal}^2$ over $200$ realizations highlighting the strong fluctuations brought in by disorder.}
        \label{fig:breakdown}
\end{figure}

The exact expression (\ref{eq:43}) shows that $\sigma_{temporal}^2$ increases as a function of bias $g$ and in fact diverges for $L_g \leq 2\xi$ (Fig. \ref{fig:vg-sec3}). The regime $\xi < L_{g} \leq 2\xi$, where the history-to-history fluctuations of transit time are anomalously large, even though $v_{drift}$ is finite, is the Anomalous Fluctuation regime. The regime $L_g<\xi$ corresponds to $v_{drift} = 0$ and defines the Vanishing Velocity regime. Fig.\ref{fig:breakdown} shows the variation of the mean transit time and the variance $\sigma^2_{temporal}$ as a function of $N$ in the normal, AF and VV regimes.

For each realization of disorder, the history-averaged mean transit time itself is different. The quantity $\sigma_{disorder}^2 = \langle \overline{T}_N^2\rangle - \langle \overline{T}_N\rangle^2$ is the contribution of different mean values to the variance. Using (\ref{eq:13}), we find 
\begin{equation}\label{eq:44}
\begin{aligned}
    \sigma_{disorder}^2 &= \frac{N e^{(2/L_{g}-1/\xi)}\big[ 1-e^{-1/\xi} \big]}{w^2\big[1- e^{(1/L_{g}-1/\xi)}\big]^2\big[1- e^{(2/L_{g}-1/\xi)}\big]}, &L_{g}>2\xi\\
    &= \infty, &L_{g}\leq 2\xi.
\end{aligned}
\end{equation}
Notice that the $\sigma_{disorder}^2$ diverges for $L_g \leq 2\xi$ as does $\sigma_{temporal}^2$. Though they are different functions of bias, their qualitative behaviour is the same (Fig. \ref{fig:vg-sec3}). A third type of variance, $\sigma_3^2 = \langle\overline{T_N^2}\rangle - \langle \overline{T_N}\rangle^2$ $ = \langle\, \overline{(T_N - \langle\overline{T_N}\rangle)^2}\,\rangle$, accounts for fluctuations of transit time about the disorder averaged mean $\langle \overline{T_N}\rangle$. We note that $\sigma_3^2 = \sigma_{temporal}^2 + \sigma_{disorder}^2$, and do not discuss it further.

Within the FTM we see that the mean drift velocity, $v_{drift}$, is a monotonically decreasing function of $g$ (Fig. \ref{fig:vg-3}) as there is no regime of diffusive motion as $g\to 0$. By making the transition rate $w$, along the backbone, a function of $g$, one may engineer a non-monotonic behaviour of $v_{drift}$ (Fig. \ref{fig:vg-4}).

The variance of transit times, $\sigma_{temporal}^2/N$, is finite for $L_g > 2\xi$ which implies that the central limit theorem holds resulting in a Gaussian distribution of transit times in the normal transport regime. For $L_g\leq 2\xi$, we enter the Anomalous Fluctuation regime where the variance diverges and the central limit theorem breaks down. The transit time distribution then follows a stable distribution with index $x = L_g/\xi$ \cite{White_1984}. Figure \ref{fig:CLT} shows that the variance grows linearly with system size $N$ in the normal transport regime, while it grows faster than linearly in the AF regime. In the Vanishing Velocity regime, i.e. $L_g \leq \xi$, the mean transit time also diverges and the law of large numbers breaks down (Fig. \ref{fig:LoLN}). The typical transit time follows a Levy stable distribution of degree $x$ with a norming constant $N^{1/x}$ in the VV regime \cite{Dhar_1984,White_1984}.
\section{Continuous and first order phase transitions in the RC} \label{sec:OZcorr}
With a purely exponential distribution of branch lengths ($\lambda = 0$ in Eq. (\ref{eq:Gammar})) we have seen that $\langle \overline{T}_N \rangle$ diverges continuously at $g_1$ (where $L_{g_1} = \xi$ from Eq. (\ref{eq:41})), and that $\sigma_{temporal}^2$ and $\sigma_{disorder}^2$ diverge continuously at $g_2$ (where $L_{g_2} = 2\xi$ from Eqs. (\ref{eq:43}) and (\ref{eq:44})) implying that the transitions to the VV and AF regimes are of the second order.

\begin{figure}[!htb]
     \centering
     \begin{subfigure}{0.45\textwidth}
         \centering
         \includegraphics[width=\textwidth]{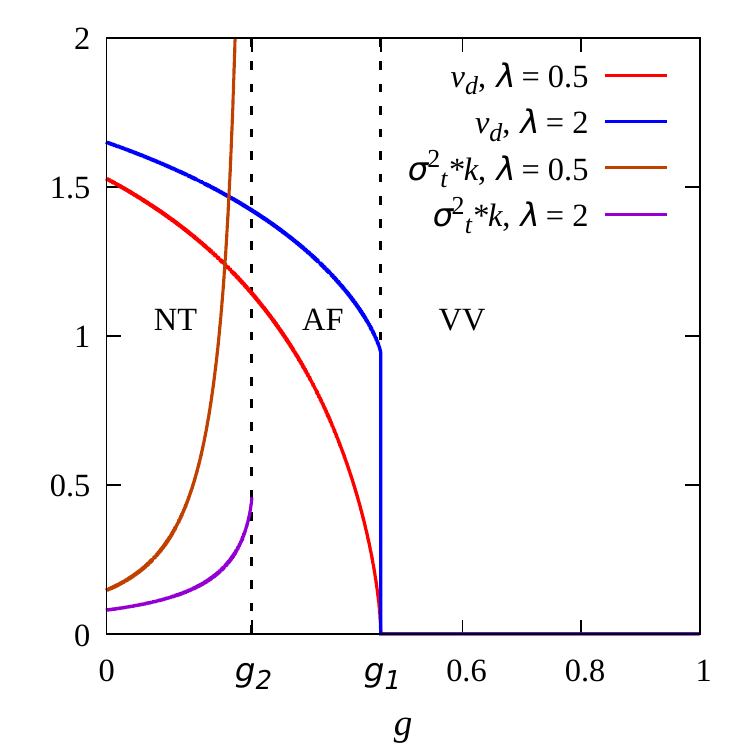}
         \caption{$w =$ constant}
         \label{fig:vg-3}
     \end{subfigure}
     \hfill
     \begin{subfigure}{0.45\textwidth}
         \centering
         \includegraphics[width=\textwidth]{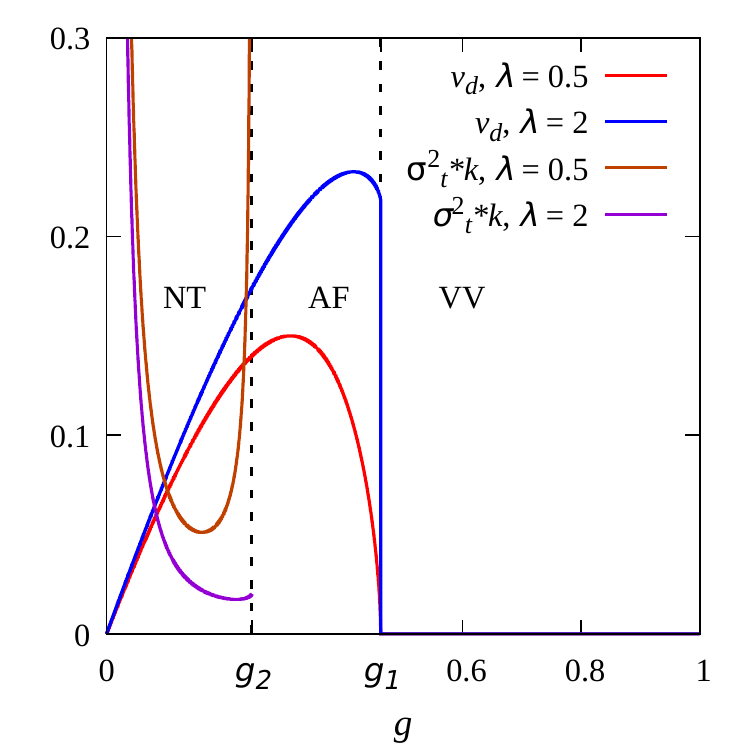}
         \caption{$w = g$}
         \label{fig:vg-4}
     \end{subfigure}
    \caption{When $\lambda>1$ ($= 2$ here), a first order phase transition is observed in $v_{drift}$ and $\sigma_{temporal}^2$. The variance $\sigma_{temporal}^2$ has been scaled down by a factor $k = 5 \times 10^{-3}$ for clarity.}
    \label{fig:firstorder}
\end{figure}

As pointed out in \cite{Demaerel_2018}, if the power law exponent $\lambda=2$, the transition at $g_1$ becomes discontinuous, with $\langle \overline{T}_N\rangle$ being finite at $g_1$. This is verified from the expression
\begin{equation}
    \langle \overline{T}_N\rangle = \frac{N}{w(e^{1/L_g}-1)}\Bigg[ \frac{e^{1/L_g}}{Z}S\bigg(\frac{1}{\xi}-\frac{1}{L_g},\lambda\bigg) - 1\Bigg]
\end{equation}
where
\begin{equation}
    S(\alpha,\lambda) = \sum_{L=0}^\infty\frac{e^{-\alpha L}}{1+(L/L_0)^{\lambda}}.
\end{equation}
At the transition point $L_{g_1} = \xi$, we see that $\langle \overline{T}_N\rangle$ diverges if $\lambda\leq 1$ and converges if $\lambda >1$. Thus, the transition is continuous for $\lambda \leq 1$ and first order for $\lambda > 1$ (Fig. \ref{fig:firstorder}). The analogous question arises about the order of the transition at $g_2$, where $L_{g_2} = 2\xi$. We have,
\begin{multline}
    \sigma_{temporal}^2 = N\Bigg[ \frac{A}{Z}\,S\bigg(\frac{1}{\xi}-\frac{2}{L_g},\lambda\bigg) - \frac{B(w+v)}{Z}S\bigg(\frac{1}{\xi}-\frac{1}{L_g},\lambda\bigg) \\
    + \frac{2Bw}{Z} S'\bigg(\frac{1}{\xi}-\frac{1}{L_g},\lambda\bigg) - C \Bigg]
\end{multline}
where $S'(\alpha,\lambda) = \partial S(\alpha,\lambda)/\partial \alpha$ and $A,B$ and $C$ were defined in Section \ref{subsec:ResT}. Setting $L_{g_2} = 2\xi$, we see that $\sigma_{temporal}^2$ diverges if $\lambda\leq 1$, implying that this transition is continuous, while $\sigma_{temporal}^2$ is finite for $\lambda>1$, implying a first order jump (Fig. \ref{fig:firstorder}). The situation is summarized succinctly in Fig. \ref{fig:phasediag-lmu}.

\section{Fluctuations in transit time in the diluted Bethe lattice} \label{sec:Bethe}
Consider biased diffusion on a randomly diluted Bethe lattice in which a certain fraction of the bonds have been removed randomly. Earlier it was shown \cite{White_1984} that beyond a certain value of the bias $g_1'$, the mean transit time along the backbone diverges, implying a vanishing drift velocity. Thus $g_1'$ is an upper bound on the critical bias $g_1$ for the transition to the VV regime. In this section, we will show that beyond a value $g_2'$, the fluctuations become anomalous. We study the steady state properties and analyze only $\sigma_{disorder}^2$. We find an upper bound $g_2'$ on the value of bias where it diverges.

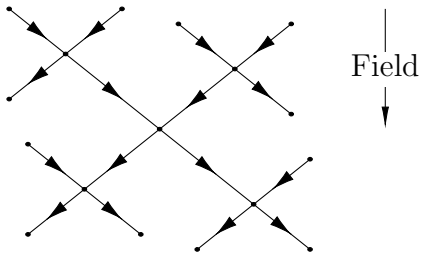
\begin{figure}[!htb]
    \centering
    \begin{tikzpicture}[yscale=0.8]
      \draw (0,0) -- (4,-4);
      \draw[fill] (0,0) circle [radius=0.03];
      \draw[fill] (0.75,-0.75) circle [radius=0.03];
      \draw[fill] (2,-2) circle [radius=0.03];
      \draw[fill] (3.25,-3.25) circle [radius=0.03];
      \draw[fill] (4,-4) circle [radius=0.03];
      
      \draw (0.75,-0.75) -- (-0.75+0.75,-0.75-0.75);
      \draw (0.75,-0.75) -- (0.75+0.75,0.75-0.75);
      \draw (2,-2) -- (1+2,1-2);
      \draw (2,-2) -- (-1+2,-1-2);
      \draw (3,-1) -- (0.75+3,0.75-1);
      \draw (1,-3) -- (-0.75+1,-0.75-3);
      \draw (3.25,-3.25) -- (-0.75+3.25,-0.75-3.25);
      \draw (3.25,-3.25) -- (0.75+3.25,0.75-3.25);
      \draw (1,-3) -- (-0.75+1,0.75-3);
      \draw (1,-3) -- (0.75+1,-0.75-3);
      \draw (3,-1) -- (-0.75+3,0.75-1);
      \draw (3,-1) -- (0.75+3,-0.75-1);

      \draw[fill] (-0.75+0.75,-0.75-0.75) circle [radius=0.03];
      \draw[fill] (0.75+0.75,0.75-0.75) circle [radius=0.03];
      \draw[fill] (-1+2,-1-2) circle [radius=0.03];
      \draw[fill] (-0.75+1,-0.75-3) circle [radius=0.03];
      \draw[fill] (1+2,1-2) circle [radius=0.03];
      \draw[fill] (0.75+3,0.75-1) circle [radius=0.03];
      \draw[fill] (-0.75+3.25,-0.75-3.25) circle [radius=0.03];
      \draw[fill] (0.75+3.25,0.75-3.25) circle [radius=0.03];
      \draw[fill] (-0.75+1,0.75-3) circle [radius=0.03];
      \draw[fill] (0.75+1,-0.75-3) circle [radius=0.03];
      \draw[fill] (-0.75+3,0.75-1) circle [radius=0.03];
      \draw[fill] (0.75+3,-0.75-1) circle [radius=0.03];
      
      \draw[-{Latex[length=3mm, width=1.5mm]}] (0.5,-0.5)--(0.51,-0.51);
      \draw[-{Latex[length=3mm, width=1.5mm]}] (1.5,-1.5)--(1.51,-1.51);
      \draw[-{Latex[length=3mm, width=1.5mm]}] (2.75,-2.75)--(2.751,-2.751);
      \draw[-{Latex[length=3mm, width=1.5mm]}] (3.75,-3.75)--(3.751,-3.751);
      \draw[-{Latex[length=3mm, width=1.5mm]}] (-0.5+0.75,-0.5-0.75)--(-0.51+0.75,-0.51-0.75);
      \draw[-{Latex[length=3mm, width=1.5mm]}] (0.251+0.75,0.251-0.75)--(0.25+0.75,0.25-0.75);
      \draw[-{Latex[length=3mm, width=1.5mm]}] (-0.7+2,-0.7-2)--(-0.71+2,-0.71-2);
      \draw[-{Latex[length=3mm, width=1.5mm]}] (0.351+2,0.351-2)--(0.35+2,0.35-2);
      \draw[-{Latex[length=3mm, width=1.5mm]}] (-0.5+1,-0.5-3)--(-0.51+1,-0.51-3);
      \draw[-{Latex[length=3mm, width=1.5mm]}] (0.251+3,0.251-1)--(0.25+3,0.25-1);
      \draw[-{Latex[length=3mm, width=1.5mm]}] (-0.5+3.25,-0.5-3.25)--(-0.51+3.25,-0.51-3.25);
      \draw[-{Latex[length=3mm, width=1.5mm]}] (0.251+3.25,0.251-3.25)--(0.25+3.25,0.25-3.25);
      \draw[-{Latex[length=3mm, width=1.5mm]}] (1.75+1,-1.75+1)--(1.751+1,-1.751+1);
      \draw[-{Latex[length=3mm, width=1.5mm]}] (1.75+1.75,-1.75+1-0.75)--(1.751+1.75,-1.751+1-0.75);
      \draw[-{Latex[length=3mm, width=1.5mm]}] (1.75+1-2,-1.75+1-2)--(1.751+1-2,-1.751+1-2);
      \draw[-{Latex[length=3mm, width=1.5mm]}] (1.75+1.75-2,-1.75+1-0.75-2)--(1.751+1.75-2,-1.751+1-0.75-2);
      \draw[-{Latex[length=3mm, width=1mm]}]  (5,0) to node [midway,fill=white, text height=1ex] {Field} (5,-2);
      \draw[white] (-1,0)--(-1,0.01);
    \end{tikzpicture}
    \caption{Easy directions in the Bethe lattice for z = 2.}
    \label{fig:Bethe}
\end{figure}

In the Bethe lattice under consideration, each site has an even coordination number, $2z$. The bonds are classified into two types based on their alignment with respect to the `easy direction' set by the field, as illustrated in Fig. \ref{fig:Bethe} for $z = 2$.

Consider diluting the structure by randomly removing a fraction of $q = (1-p)$ of all bonds. Above the percolation threshold, $p_c = 1/(2z-1)$, infinite clusters form in which each site is connected to infinity through at least one path. Sites that have at least two ways to connect to infinity constitute the backbone. Finite clusters connected at a single site to the backbone are called branches.

In order to find the total transit time along the backbone in the steady state, we use the fact that $\overline{T} = \mathcal{N}/J$, where $\mathcal{N}$ is total number of particles and $J$ is the current injected into the backbone in the steady state. Let $\rho_n$ be the density of particles at a backbone site $n$. The density at a branch site $m$ is related to $\rho_n$ through
\begin{equation}
    \rho_m = e^{-r_{nm}/L_g}\rho_n
\end{equation}
where $r_{nm}$ is the difference between total number of up bonds and down bonds connecting $n$ to $m$. The total number of particles in a branch $\beta$, associated with site $n$ is then just $\rho_n$ times a factor $F_{n\beta}$, given by
\begin{equation}
    F_{n\beta} = \sum_{m\in\beta}e^{-r_{nm}/L_g}.
\end{equation}
The total number of possible configurations of a branch depend only on whether it is an up branch ($r_{n1} = +1$) or a down branch ($r_{n1} = -1$). The average of $F_{\pm}$ over all the up or down branches is given by
\begin{equation}\label{eq:F_av}
    \langle F_{\pm}\rangle = \sum_{m}e^{-r_{nm}/L_g}\langle\eta_{nm}\rangle.
\end{equation}
The quantity $\eta_{nm}$ is an indicator function such that $\eta_{nm} = 1$ if site $m$ belongs to branch $\beta$ and $0$ otherwise. Taking an average over all the possible configurations of $\beta$, we obtain
\begin{equation}
\begin{aligned}
    \langle F_{\pm}\rangle &= p\big[\mathcal{Q}(p)\big]^{2z-1}\sum_{k=1}^2\frac{C_k^{\pm}(g)}{1-\lambda_k(g)p\big[\mathcal{Q}(p)\big]^{2z-2}}, &g<g_1',\\
    &=\infty, &g\geq g_1'.
\end{aligned}
\end{equation}
Details of derivation are given in \ref{sec:appendixB}, here $g_0$ is the solution of the equation $\lambda_1(g)p\big[\mathcal{Q}(p)\big]^{2z-2} = 1$, and the functions $\lambda_k(g), \mathcal{Q}(p)$ and $C_k^{\pm}(g)$ are defined in \ref{sec:appendixB}.

The evaluation of $\sigma_{disorder}^2$ involves calculating
\begin{equation}
    \Delta F_{\pm}^2 = \langle F_{\pm}^2\rangle - \langle F_{\pm}\rangle^2.
\end{equation}
The first term on the right is given by
\begin{equation}\label{eq:F2}
    \langle F_{\pm}^2\rangle = \sum_{m}e^{-2r_{nm}/L_g}\langle\eta_{nm}\rangle + \sum_{m\neq m'}e^{-(r_{nm}+r_{nm'})/L_g}\langle\eta_{nm}\eta_{nm'}\rangle.
\end{equation}
From Eqs. (\ref{eq:F_av}) and (\ref{eq:F2}) and using the independence of $\eta_{nm}$ and $\eta_{nm'}$, we obtain
\begin{equation}\label{eq:51}
    \Delta F_{\pm}^2 = \sum_{m}e^{-2r_{nm}/L_g}\langle\eta_{nm}\rangle(1-\langle\eta_{nm}\rangle).
\end{equation}
The final result for $\Delta F_{\pm}^2$ is
\begin{equation}
\begin{aligned}
    \Delta F_{\pm}^2 =& p\big[\mathcal{Q}(p)\big]^{2z-1}\sum_{k=1}^2\frac{D_k^{\pm}}{1-\alpha_k(g)p\big[\mathcal{Q}(p)\big]^{2z-2}} \\
    &- p^2\big[\mathcal{Q}(p)\big]^{4z-2}\sum_{k=1}^2\frac{D_k^{\pm}}{1-\alpha_k(g)p^2\big[\mathcal{Q}(p)\big]^{4z-4}}, &\textrm{if}\, g<g_2',\\
    =& \infty, &\textrm{if}\, g\geq g_2'
\end{aligned}
\end{equation}
where $g_2'$ is the solution of the equation $\alpha_1(g)p\big[\mathcal{Q}(p)\big]^{2z-2} = 1$. As shown in the \ref{sec:appendixB}, $\alpha_k(L_g) = \lambda_k(L_g/2)$ and $D_k^{\pm}(L_g) = C_k^{\pm}(L_g/2)$, which implies that $L_{g_2'} = 2L_{g_1'}$ and hence $g_2'<g_1'$. Therefore, the variance of mean transit times diverges at a lower value of bias than the mean transit time itself, pointing to the existence of a distinct regime of anomalous fluctuations in the dilute Bethe lattice.

In the analysis above, we have accounted only for trapping by branches. There is the possibility of trapping by valleys along the backbone, which means that $g_2'$ is, in principle, an upper bound to the true threshold for the AF regime.
\section{Conclusions} \label{sec:discussion}
We may draw two broad conclusions from our study of field-induced drift and trapping for biased random walkers on the random comb: first, that there is a separate anomalous fluctuation (AF) regime between the normal transport and vanishing velocity (VV) regimes, as indicated by a divergence of the exactly computed variance of transit times; and secondly, that the transition to the AF regime changes from continuous to first order if the power law corrections to the exponential form (Eq. (\ref{eq:Gammar})) are strong enough. These conclusions are summarized in the phase diagram in Fig. \ref{fig:phasediag-lmu}.

In the regime where the variance is finite, we calculated $\sigma_{temporal}^2$ and $\sigma_{disorder}^2$ exactly, in the forward transport limit on the random comb. In this limit, the transit time is simply the sum of the residence times on each backbone site. In order to ensure a uniform transition rate in each bond of the backbone, we showed that rates for random walk motion need to be site-dependent. Uniformity of backbone-bond rates implies that the steady state particle density is uniform along the backbone and rises exponentially in the branches.

The distribution of first return times in a branch, which is a prerequisite to finding the distribution of residence times, itself shows three distinct behaviours, corresponding to three smoothly connected regimes: diffusive, with a distribution $\sim 1/t^{3/2}$; drift, associated with a rapidly decaying exponential distribution; and a trapping regime with a distribution of ultra-slow return times which grow exponentially with branch length, reflecting the difficulty of crossing the bias-induced barrier. These points are brought out clearly by numerical simulations, as shown in Fig. \ref{fig:frt}. A non-standard procedure, described in \ref{sec:appendixA}, had to be followed for the numerical simulations in view of the extremely large times involved. 

The occurrence of the vanishing velocity regime, known from earlier work \cite{White_1984,POTTIER19951,BALAKRISHNAN19951} is associated with the breakdown of the law of large numbers, as the mean residence time on a site becomes infinite. Likewise, the AF regime is associated with the breakdown of the central limit theorem, owing to a diverging variance of the residence time distribution. In \cite{Dhar_1984}, it was argued that the time-distance relation in the VV regime asymptotically follows $T\sim R^{1/x}$ with $x=L_g/\xi$. In the AF regime, we expect that asymptotically, $T\sim R/v_{drift}$, but $\Delta T^2 \sim R^{2/x}$.

The order of the transitions is sensitive to the distribution of branch lengths for the random comb. It becomes first order at the bias thresholds of both the AF and VV regimes once $\lambda$ exceeds $1$.

Finally, we obtained an upper bound on the critical bias for the variance $\sigma_{disorder}^2$ to diverge on the Bethe lattice with bond dilution. This indicates the presence of an anomalous fluctuation regime for this structure as well.

We conclude by pointing to interesting open problems involving other types of stochastically moving particles on the random comb. Two examples: (a) interacting particles with mutual exclusion and (b) non-interacting active particles, undergoing run-and-tumble motion. Problem (a) has been studied earlier \cite{Ramaswamy_1987,Demaerel_2018} and it has been argued that there is no VV regime, as branches essentially fill up and their lengths are screened, thereby limiting mean times spent in a branch. It would be interesting to carry out a study of fluctuations in this system, to see if there is an AF regime. For case (b) one needs to ascertain whether there is a VV regime at all, and even if not, whether there is an AF regime.
\section*{Acknowledgements}
This work forms part of a thesis submitted by J.D.K. in partial
fulfilment of the requirement for the degree of Master of
Science awarded by Indian Institute of Space Science and
Technology, Thiruvananthapuram. We acknowledge useful discussions with Arghya Das, Kabir Ramola, Stephy Jose and Navdeep Rana. J.D.K. would like to thank TIFR Hyderabad for hospitality and IIST for academic support. M.B. acknowledges support under the DAE Homi Bhabha Chair Professorship of the Department of Atomic Energy, India. This project was funded by intramural funds at TIFR Hyderabad from the Department of Atomic Energy (DAE), Government of India.
\appendix
\section{Numerical procedure: Simulating random walks on random combs} \label{sec:appendixA}
We perform numerical simulations to get an insight into random walk dynamics on random combs. The complete procedure is divided into three steps as follows.
\renewcommand{\thesubsection}{\arabic{subsection}}
\subsection{Generating realizations of disorder}\label{subsec:Realz}
We generate and store multiple different realizations of disorder randomly from the distribution (\ref{eq:Gammar}). Each realization contains $9000$ branches and we generate $1000$ such realizations. The maximum branch length, $L_{max}$, across the realizations is recorded. With the value $\xi = 1$, we found $L_{max} = 16$.

\subsection{First return times in branches: Setting up a Library}
A library contains a collection of first return times for branch lengths $L = 1,2,\dotsc,L_{max}$ for $10^6$ different histories. The first return times on a given branch length, corresponding to different histories, are stored sequentially in a single column. Each column corresponds to a different branch length $L$. Multiple libraries are generated, each with a different value of $g$.
    
The first return times for a branch of length $L$ with bias $g$ are generated as follows. Let the particle be at $m=0$, at $t=0$. We draw a time randomly from the distribution $\phi(t)$, with $\gamma = u+w$, and update the particle position to $m = 1$. Again a time $t$ is randomly drawn from $\phi(t)$, but with $\gamma = u+v$ this time, and the position updated to $m \pm 1$ with a probability $p$ or $q$ (Section \ref{sec:RC}). The process is continued until $m = 0$ again or if $t$ crosses a preset value $t_{max}$. In the later case, the time is stored as $t_{max}$. The total time taken for the excursion is stored and the process is repeated for $10^6$ histories.
    
In case a time $t_{max}$ is drawn from the library (for computing the residence time), we draw another time randomly from the normalized distribution $1/\tau_L\exp(-t/\tau_L)$ and add it to $t_{max}$. The value of $t_{max}$ is chosen to be large enough to always be in the trapping regime, where the times are governed by $\sim A\exp(-t/\tau_L)$, $A$ being some constant.
    
The time $\tau_L$ is determined by fitting an exponential curve to the numerically obtained profiles of first-return times (Fig. \ref{fig:frt}). We find that $\tau_L = C(g)e^{L/L_g}$, where $C(g)$ is a constant for a given $g$. Fitting the curve of $\tau_L$ vs $L$ to an exponential we determine the value of $C(g)$. This curve is extrapolated to determine $\tau_L$ for large $L$.

\subsection{Total transit time: Summing the residence times}
As it progresses along the backbone, the typical trajectory of a particle on the RC involves performing multiple excursions into a branch, waiting at the corresponding backbone site, hopping to the next backbone site and repeating it for $N$ backbone sites. In a particular history, the number of excursions $n$ at a given site is chosen from the distribution $r^n(1-r)$ (Fig. \ref{fig:RCrates}) and $n$ first return times are drawn from the library. These times are summed up and a waiting time, drawn from $\phi(t)=(u+w)\exp(-(u+w)t)$, is added to the sum to obtain the residence time.
    
In a given realization of disorder, the residence time is computed for all the branches and added to obtain the total transit time. We perform $10^4$ histories over a given realization of disorder and average over $10^3$ realizations of disorder. The procedure is repeated for each of the $13$ different values of $g$.

\section{Calculations on the Bethe lattice} \label{sec:appendixB}

Here we evaluate the sum (\ref{eq:51}) using the method proposed in \cite{White_1984}. The quantity $\eta_{nm}$ is the indicator function, the average of which is given by
\begin{equation}
    \langle \eta_{nm}\rangle = p^N\big[\mathcal{Q}(p)\big]^{(2z-2)N+1}
\end{equation}
where $\mathcal{Q}(p)$ is the probability that all the paths in a given direction from an occupied site are finite and $N$ is the total number of bonds between $n$ and $m$, which follows
\begin{equation}
    \mathcal{Q}(p) = q + p\big[\mathcal{Q}(p)\big]^{2z-1}.
\end{equation}
For simplicity, let us assign each bond, a pseudospin $s_i$ such that $s_i = +1$ for an `up' bond and $s_i = -1$ for a `down' bond. Therefore, $N = \sum_{k\in\beta}|s_k|$ and $r_{nm} = \sum_k s_k$. At every branch site $m+1$ there are $z$ ways of having $s_{m+1} = s_m$ and $z-1$ ways of having $s_{m+1} = -s_m$. Let $n(s_{m+1},s_m)$ be such that
\begin{equation}
\begin{split}
    n(s_{m+1},s_m) &= z, \qquad\quad\quad s_{m+1} = s_m, \\
    &= z-1, \qquad\, s_{m+1} = - s_m.
\end{split}
\end{equation}
It is found that
\begin{equation}\label{eq:B1}
    \langle F_{s_1}\rangle = \mathcal{Q}(p)\sum_{N=1}^{\infty}\nu^N(C_1^{\pm}\lambda_1^{N-1}+C_2^{\pm}\lambda_2^{N-1})
\end{equation}
where $\lambda_1(g)$ and $\lambda_2(g)$ are the eigenvalues of the transfer matrix $T_1$
\begin{equation}\label{eq:B3}
    T_1 = 
\begin{pmatrix}
z\,e^{-h} & z-1 \\
z-1 & z\,e^{h} 
\end{pmatrix}
\end{equation}
with $\lambda_1>\lambda_2$,
\begin{equation}
    \lambda_k(g) = z\cosh h \pm \big[z^2\sinh^2h+(z-1)^2\big]^{1/2}
\end{equation}
and the coefficients $C_1^{\pm}$ and $C_2^{\pm}$ are given by
\begin{equation}
    C_{k}^{\pm} = \frac{(z-1)(\lambda_k-e^{\mp h})}{2(\lambda_k - z\cosh h)(\lambda_k - z\,e^{\mp h})}
\end{equation}
where we have substituted $h = 1/L_g$ and $\nu = p\big[\mathcal{Q}(p)\big]^{(2z-2)}$. The sum  in (\ref{eq:B1}) converges for $\lambda_1(g)\,p\big[\mathcal{Q}(p)\big]^{2z-2}<1$, and is given by
\begin{equation}
    \langle F_{\pm}\rangle = p\big[\mathcal{Q}(p)\big]^{2z-1}\sum_{k=1}^{2}\frac{C_k^{\pm}(g)}{1-\lambda_k(g)p\big[\mathcal{Q}(p)\big]^{2z-2}}.
\end{equation}
Let $g_1'$ be the solution of $\lambda_1(g)\,p\big[\mathcal{Q}(p)\big]^{2z-2} = 1$. The variance of transit time is given by,
\begin{equation}
\begin{aligned}\label{eq:B2}
    \Delta F_{\pm}^2 &= \sum_m e^{-2r_{nm}/L_g}\langle\eta_{nm}\rangle\big(1-\langle\eta_{nm}\rangle\big)\\
    &= \sum_{N=1}^{\infty}\mathcal{Q}\nu^N\big(1-\mathcal{Q}\nu^N\big)\sum_{\{s_2,..,s_N\}}\prod_{k=1}^{N-1}n(s_{k+1},s_k)e^{2(s_1+\dotsc+s_N)/L_g}.
\end{aligned}
\end{equation}
Using transfer matrix $T_2$, obtained by replacing $h\to 2h$ in $T_1$, the above reduces to
\begin{equation}
    \Delta F_{\pm}^2 = \sum_{N=1}^{\infty}\mathcal{Q}(p)\nu^N\big(1-\mathcal{Q}(p)\nu^N\big)\sum_{s_N = \pm 1}e^{-(s_1+s_N)h}\langle s_1|T_2^{N-1}|s_N\rangle.
\end{equation}
Therefore,
\begin{equation}
    \Delta F_{\pm}^2 = \sum_{N=1}^{\infty}\mathcal{Q}(p)\nu^N\big(1-\mathcal{Q}(p)\nu^N\big)(D_1^{\pm}\alpha_1^{N-1}+D_2^{\pm}\alpha_{2}^{N-1})
\end{equation}
where $\alpha_1(g) > \alpha_2(g)$ are the eigenvalues of $T_2$,
\begin{equation}
    \alpha_k(g) = z\cosh 2h \pm \big[z^2\sinh^2 2h+(z-1)^2\big]^{1/2}
\end{equation}
and the coefficients $D_1^{\pm}$ and $D_2^{\pm}$ are given by
\begin{equation}
    D_{k}^{\pm} = \frac{(z-1)(\alpha_k-e^{\mp 2h})}{2(\alpha_k - z\cosh 2h)(\alpha_k - z\,e^{\mp 2h})}.
\end{equation}
The series is convergent for $\alpha_1(g)p\big[Q(p)\big]^{2z-2}<1$. Let $g_2'$ be the solution for the latter with an equality sign, thus for $g<g_2'$,
\begin{equation}
\begin{aligned}
    \Delta F_{\pm}^2 =& p\big[\mathcal{Q}(p)\big]^{2z-1}\sum_{k=1}^2\frac{D_k^{\pm}}{1-\alpha_k(g)p\big[\mathcal{Q}(p)\big]^{2z-2}} \\
    &- p^2\big[\mathcal{Q}(p)\big]^{4z-2}\sum_{k=1}^2\frac{D_k^{\pm}}{1-\alpha_k(g)p^2\big[\mathcal{Q}(p)\big]^{4z-4}}.
\end{aligned}
\end{equation}
Clearly, $\alpha_k(h) = \lambda_k(2h)$, hence, $\alpha_k(L_g) = \lambda_k(L_g/2)$. Therefore, the value of $L_g$ obtained in this case will be twice that obtained in the former case and hence $g_2'<g_1'$.
\bibliographystyle{elsarticle-num-names} 
 \bibliography{cas-refs}




\end{document}